# Day Level Forecasting for Coronavirus Disease (COVID-19) Spread: Analysis, Modeling and Recommendations

***Haytham H. Elmousalami*[2] *and Aboul Ella Hassanien*[1,2]**

[1]Cairo University, Faculty of Computers and Artificial Intelligence, Cairo, Egypt
[2]Scientific Research Group in Egypt (SRGE), Cairo, Egypt http://www.egyptscience.net
E-mail :Haythamelmousalami2014@gmail.com & aboitcairo@cu.edu.eg

**Abstract:** In mid of March 2020, Coronaviruses such as COVID-19 is declared as an international epidemic. More than 125000 confirmed cases and 4,607 death cases have been recorded around more than 118 countries. Unfortunately, a coronavirus vaccine is expected to take at least 18 months if it works at all. Moreover, COVID -19 epidemics can mutate into a more aggressive form. Day level information about the COVID -19 spread is crucial to measure the behavior of this new virus globally. Therefore, this study presents a comparison of day level forecasting models on COVID-19 affected cases using time series models and mathematical formulation. The forecasting models and data strongly suggest that the number of coronavirus cases grows exponentially in countries that do not mandate quarantines, restrictions on travel and public gatherings, and closing of schools, universities, and workplaces ("Social Distancing").

## 1. Introduction

**Coronaviruses** such as SARS, MERS, and COVID-19 are a group of viruses that infects both mammals and birds. Coronaviruses cause infections that are typically such as some cases of the common cold in humans where Symptoms vary based on the infected species (Menachery et al. 2015; Zhu et al. 2020; Huang et al. 2020). The COVID-19 has reported being a novel coronavirus of a typical pneumonia since Dec 31, 2019. The COVID-19 started in Wuhan Chinese city. Simultaneously, Both the international world and other Chinese cities have received infected cases. On 11 March 2020, the world health organization (WHO) stated that the global COVID-19 outbreak is a pandemic because of the speed and scale of transmission of the virus. From 118 countries and territories, there are 125,000 cases reported to WHO. Moreover, the number of cases announced outside China has almost doubled 13-folds in the past two weeks and the affected countries' number has almost tripled (Li et al. 2020; Wu et al. 2020). Fig.1 shows the Coronavirus COVID-19 global cases by the center for systems science and engineering (CSSE) at Johns Hopkins University (Dong and Gardner 2020).

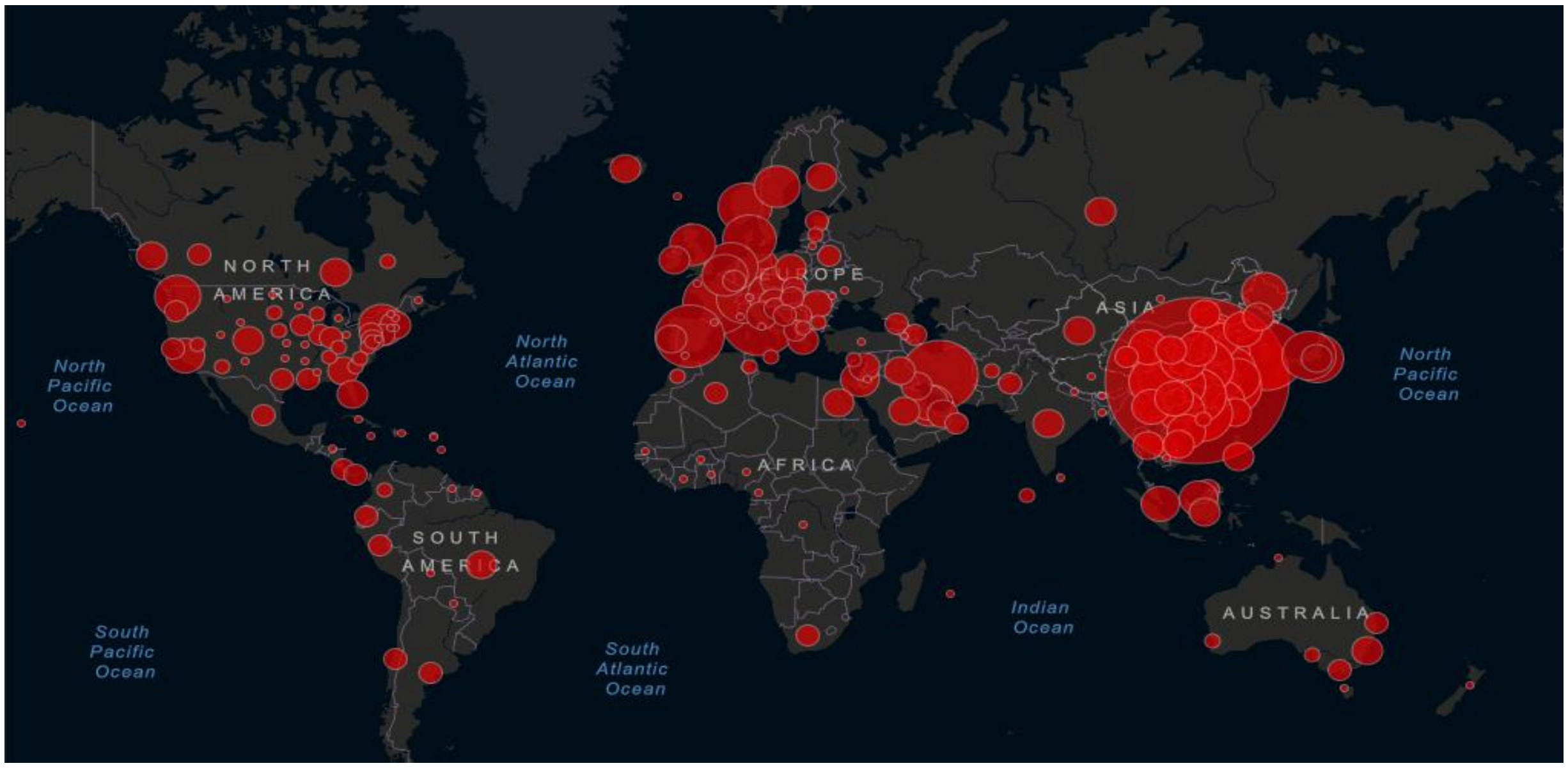

Fig.1 Coronavirus COVID-19 Global Cases (JHU, 2020).

## 2. Methods and Materials

### *2.1 Data description*

Daily level information on the affected people can give insights about the international spread of the COVID-19. World health organization (WHO), the national health commission of China and Johns Hopkins University have developed an open database for the COVID-19 cases. Therefore, the research objectives are getting insights such as the changes in several affected cases over time and forecasting the latest number of affected cases (confirmed cases, recovered cases, or deaths). Moreover, this research aims to analyze the change in cases of overtime at the country level. The data consists of the date of the observation in MM/DD/YYYY, Province or state of the observation, Country of observation, Cumulative number of confirmed cases till that date, Cumulative number of deaths till that date, Cumulative number of recovered cases till that date.

### *2.2 Time series forecasting*

Forecasting is making predictions of the future using past and present data. The major category of forecasting is qualitative methods, explanatory (causal) techniques and time series algorithms. A time series is a sequence of discrete-time data points listed in time order **(**Lin et al. 2003). Several models of time series exist such as moving average (MA), weighted moving average (WMA), and single exponential smoothing (SES). A moving average (MA) is analyzing the data points by averaging the series of data points. A moving average (MA) depends on the assumption of future observation is similar to recently previous observations. Moving average of the next point equals the average of the recent $\boldsymbol{K}$ observation. As $\boldsymbol{K}$ increases, most forecast relies on older data. Mathematically, $\boldsymbol{MA}$ can be computed in the following formula (Box et al. 1970; Yang et al. 2018):

$$\boldsymbol{MA = \frac{D_1 + D_2 + \ldots + D_k}{n}} \qquad \textbf{(1)}$$

*Where MA, D, K, and n are moving average, the observed data value, number of pints period, and number of data points, respectively.*

Similar to Moving average (MA), weighted moving average (WMA) is a modification of Moving average (MA) model by assigning weights to data points as follows (Hunter 1986; Lowry et al. 1992):

$$\mathbf{WMA} = \mathbf{D_1} * \mathbf{W_1} + \mathbf{D_2} * \mathbf{W_2} + \ldots . + \mathbf{D_K} * \mathbf{W_K} \qquad \textbf{(2)}$$

*Where WMA, D, W, K, and n are weighted moving average, the observed data value, weights, number of pints period, and number of data points, respectively.*

Single Exponential Smoothing (SES) is a smoothing time series data based on the exponential window function (Nazim and Afthanorhan 2014; Cadenas et al. 2010). Moreover, triple exponential smoothing (Holt-Winters method) is an algorithm used to forecast data points in a series. Mathematically, SES can be computed in the following formula.

$$\mathbf{F_{t+1}} = (\mathbf{1} - \boldsymbol{\alpha})\mathbf{F_t} + \boldsymbol{\alpha}\mathbf{D_t} \qquad \textbf{(3)}$$

*Where*

$\boldsymbol{F_{t+1}}$ *is the forecast for period t+1*

$\boldsymbol{F_t}$ *is the forecast for period t*

$\boldsymbol{D_t}$ *the observed data value at time = t*

$\boldsymbol{\alpha}$ *is the smoothing constant ranging from 0 to 1*

## 3. Results and analysis

The forecasting models have been conducted to predict the cumulative international confirmed, recovery and death of the COVID-19 cases. First, according to forecasting confirmed cases, Fig.2 shows the moving average (MA) model for forecasting confirmed cases using 5 weeks ahead and 10 weeks ahead scales. Fig.3 weighted moving average (WMA) for forecasting confirmed cases where the used weights are 0.0%, 0.0%, 1.5%, 38.1%, 60.3%. Fig.4 Single Exponential Smoothing (SES) for forecasting confirmed cases. Alpha is optimized based on a genetic algorithm to be equal to 1.00. The three forecasting models have successfully fit the recorded data of the confirmed cases based on different criteria such as mean absolute deviation (MAD), mean square error (MSE), root mean square error (RMSE), mean absolute percentage error (MAPE). Mathematically, these criteria can be represented using the following equations (4-6).

$$\mathbf{MAD} = \frac{\mathbf{1}}{\boldsymbol{n}}\sum_{\boldsymbol{i}=\mathbf{1}}^{\boldsymbol{n}}[\boldsymbol{y_i} - \boldsymbol{\hat{y}_i}]^{\mathbf{1}} \qquad (4)$$

Where *n* is the total summation of cases, (i) is the number of the case and $\boldsymbol{\hat{y}_i}$ is the predicted outcome of the time series model and $\boldsymbol{y_i}$ is the actual outcome.

$$\boldsymbol{MSE} = \frac{\mathbf{1}}{\boldsymbol{n}}\sum_{\boldsymbol{i}=\mathbf{1}}^{\boldsymbol{n}}[\boldsymbol{y_i} - \boldsymbol{\hat{y}_i}]^{\mathbf{2}} \qquad \textbf{(5)}$$

Where $\boldsymbol{y_i}$ are observed values, $\boldsymbol{\hat{y}_i}$ is the predicted values, and $\boldsymbol{n}$ is the number of observations.

$$(\boldsymbol{MAPE}) = \left(\frac{\boldsymbol{1}}{\boldsymbol{n}}\sum_{i=1}^{n}\frac{|y_i - \hat{y}_i|}{\hat{y}_i}\boldsymbol{x100}\right) \qquad \textbf{(6)}$$

Where *n* is the total number of cases, *i* is the number of the case and $\boldsymbol{\hat{y}_i}$ is the predicted outcome of the model and $\boldsymbol{y_i}$ is the actual outcome. As shown in Table.1, the results of forecasting models for confirmed cases are ranging from 21.42% to on 9.68% MAPE scale. The results indicate that Single Exponential Smoothing (SES) is the most accurate model for forecasting confirmed cases COVID-19 with 3385.65, 20050014.56, 4477.72, and 9.68% for mean absolute deviation (MAD), mean square error (MSE), root mean square error (RMSE), means absolute percentage error (MAPE), respectively.

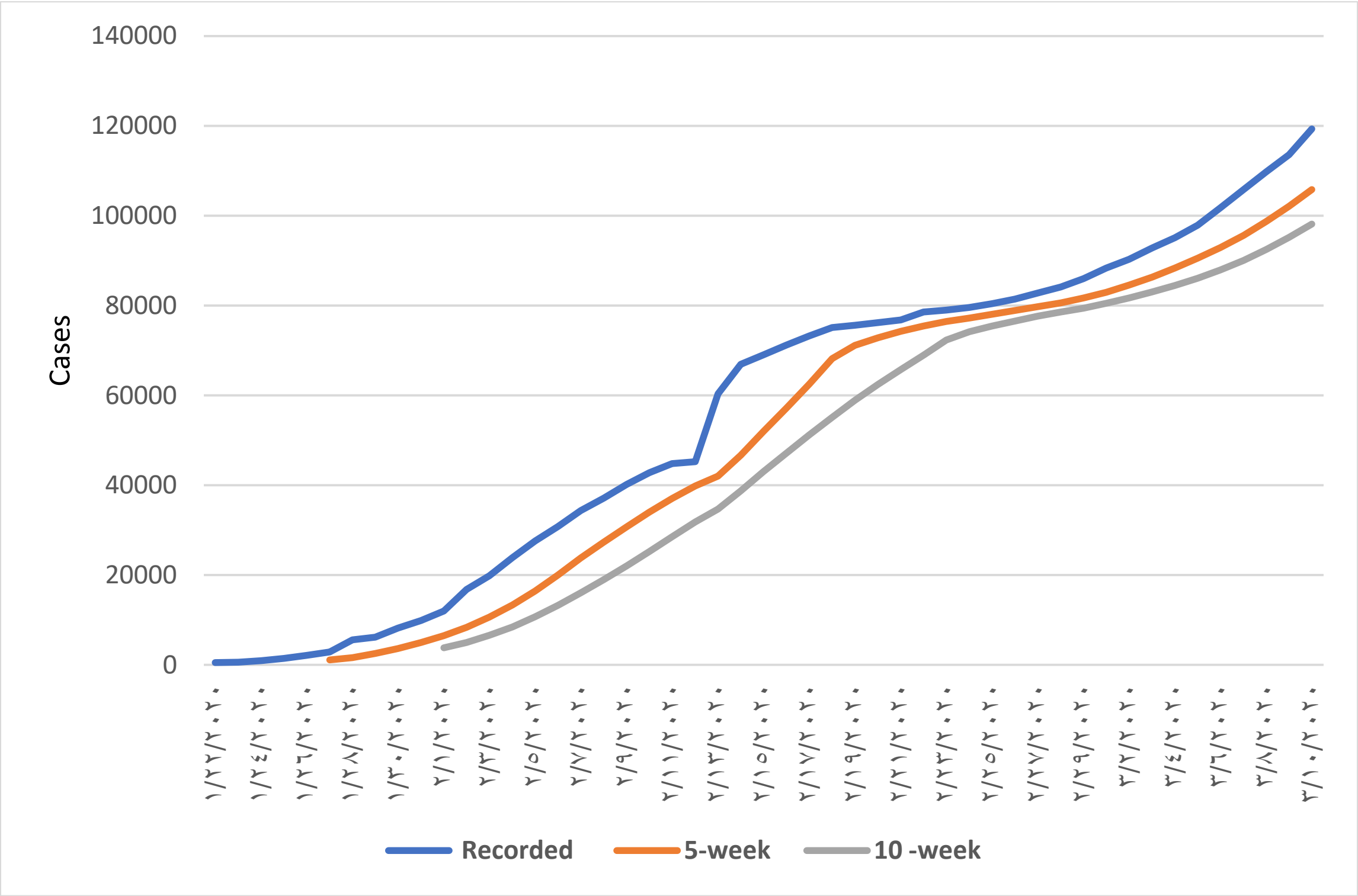

Fig.2 Moving average (MA) for forecasting international confirmed cases.

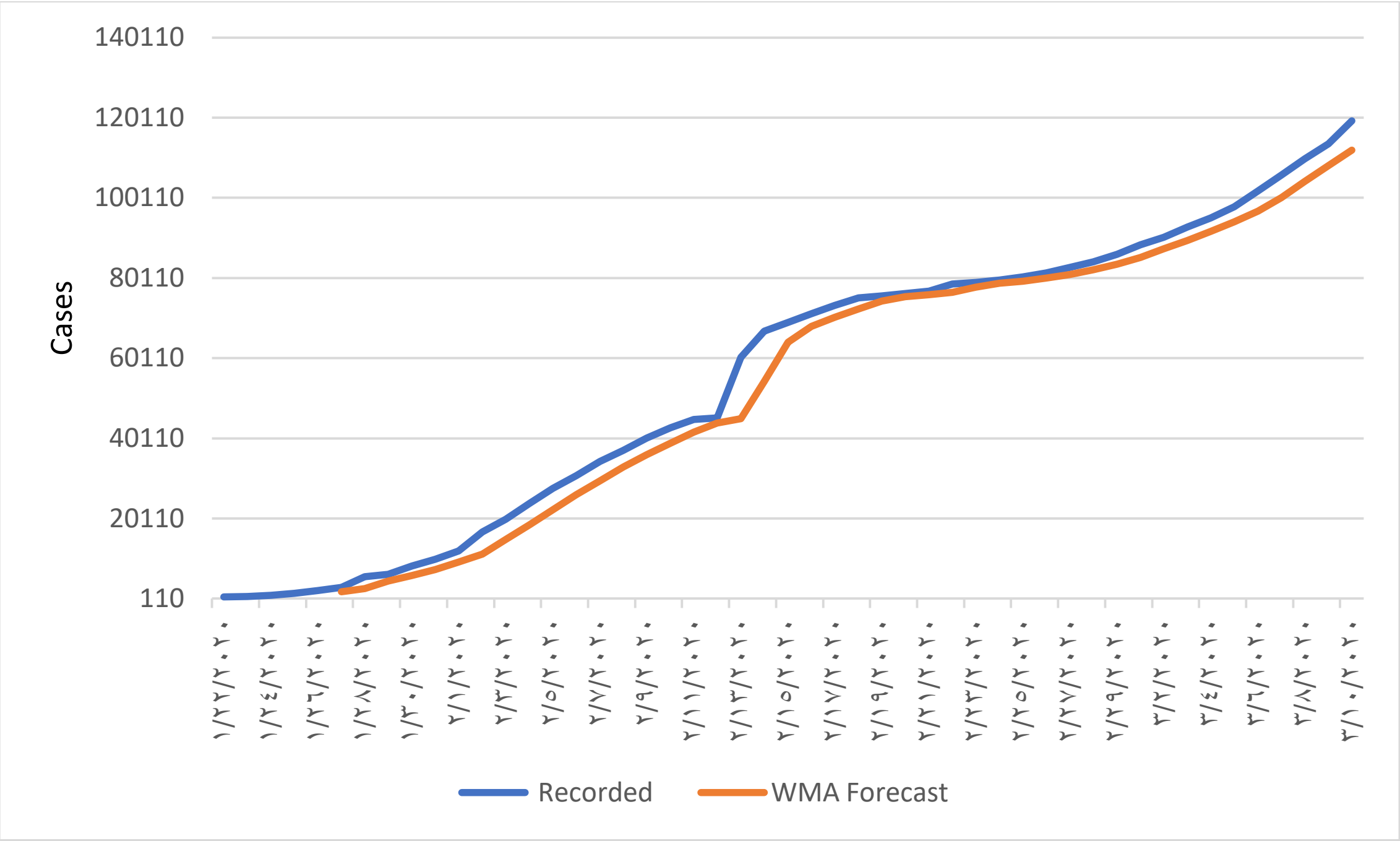

Fig.3 Weighted Moving average (WMA) for forecasting international confirmed cases.

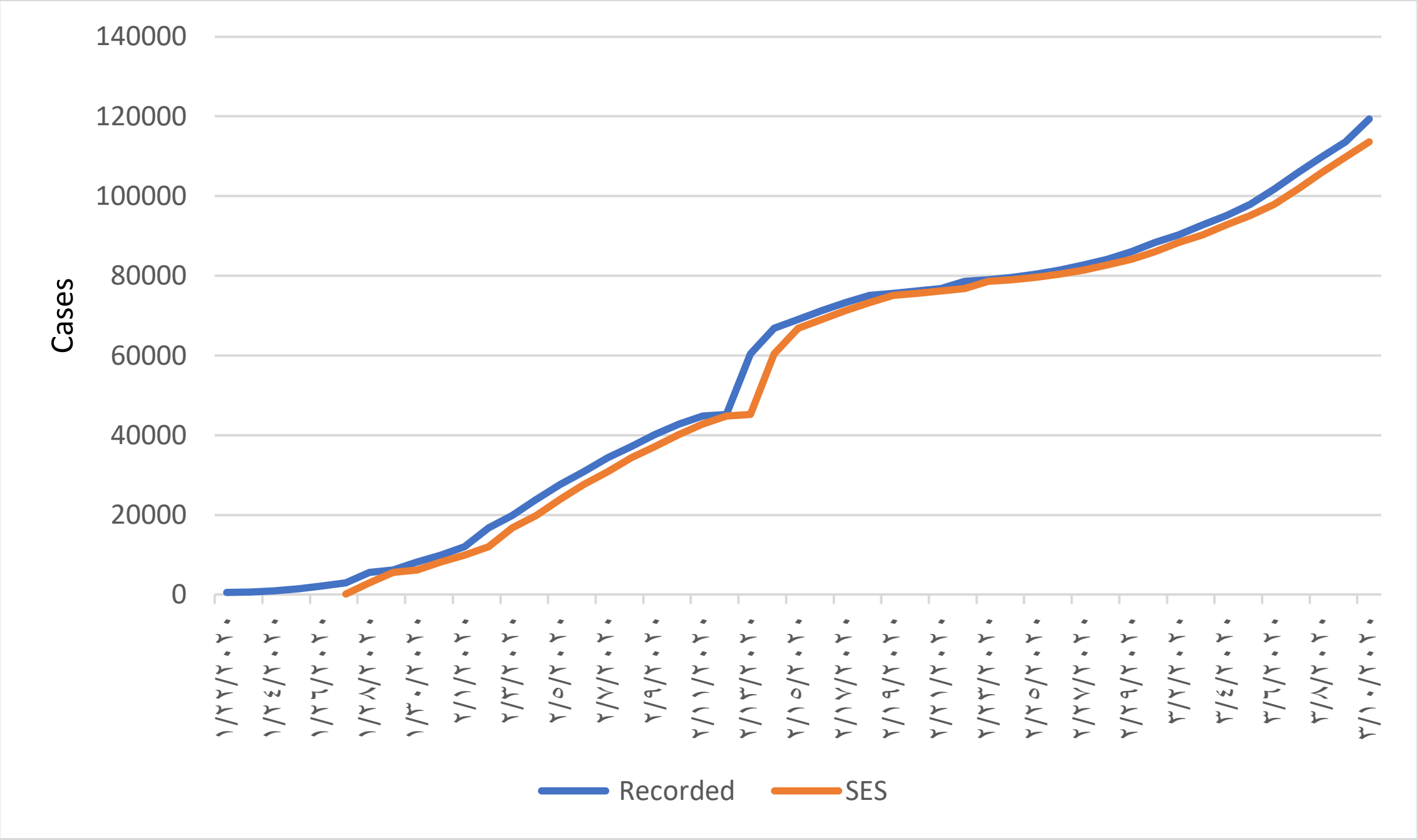

Fig.4 Single Exponential Smoothing (SES) for forecasting international confirmed cases.

Table.1 Results of forecasting models for international confirmed cases.

| Model / Criteria | MAD | MSE | RMSE | MAPE |
|---|---|---|---|---|
| **A moving average (MA)** | | | | |
| **5 weeks ahead** | 7602.60 | 77470882.22 | 8801.75 | 21.42% |
| **10 weeks ahead** | 18039.43 | 242692414.14 | 15578.59 | 28.18% |
| **Weighted Moving average (WMA)** | 4614.96 | 32413677.85 | 5693.30 | 11.16% |
| **Single Exponential Smoothing (SES)** | 3385.65 | 20050014.56 | 4477.72 | 9.68% |

Second, according to forecasting recovered cases, Fig.5 shows the moving average (MA) model for forecasting recovered cases for 5 weeks ahead and 10 weeks ahead scales. Fig.6 weighted moving average (WMA) for forecasting confirmed cases where the used weights are 0.0%, 0.0%, 1.5%, 38.1%, 60.3%. Fig.7 Single Exponential Smoothing (SES) for forecasting recovered cases. Alpha is optimized based on a genetic algorithm to be equal to 1.00. The three forecasting models have successfully fit the recorded data of the confirmed data. As shown in Table.2, the results of forecasting models for recovered cases are ranging from 34.88% to on 16.38% MAPE scale. The results indicate that Single Exponential Smoothing (SES) is the most accurate model for forecasting recovered cases COVID-19 with 517.54, 523335.16, 723.42, and 16.38% for mean absolute deviation (MAD), mean square error (MSE), root mean square error (RMSE), mean absolute percentage error (MAPE), respectively.

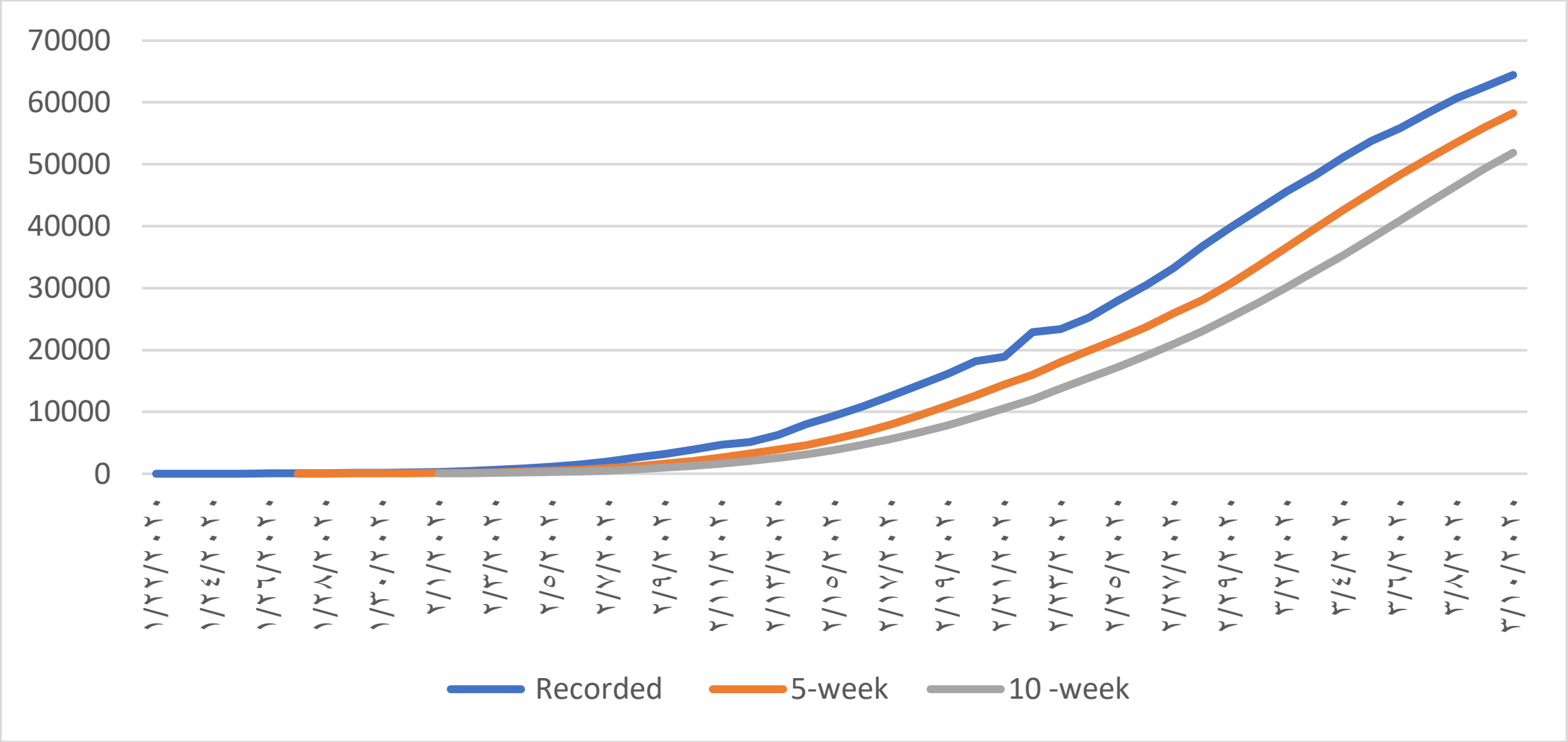

Fig.5 Moving average (MA) for forecasting international recovered cases.

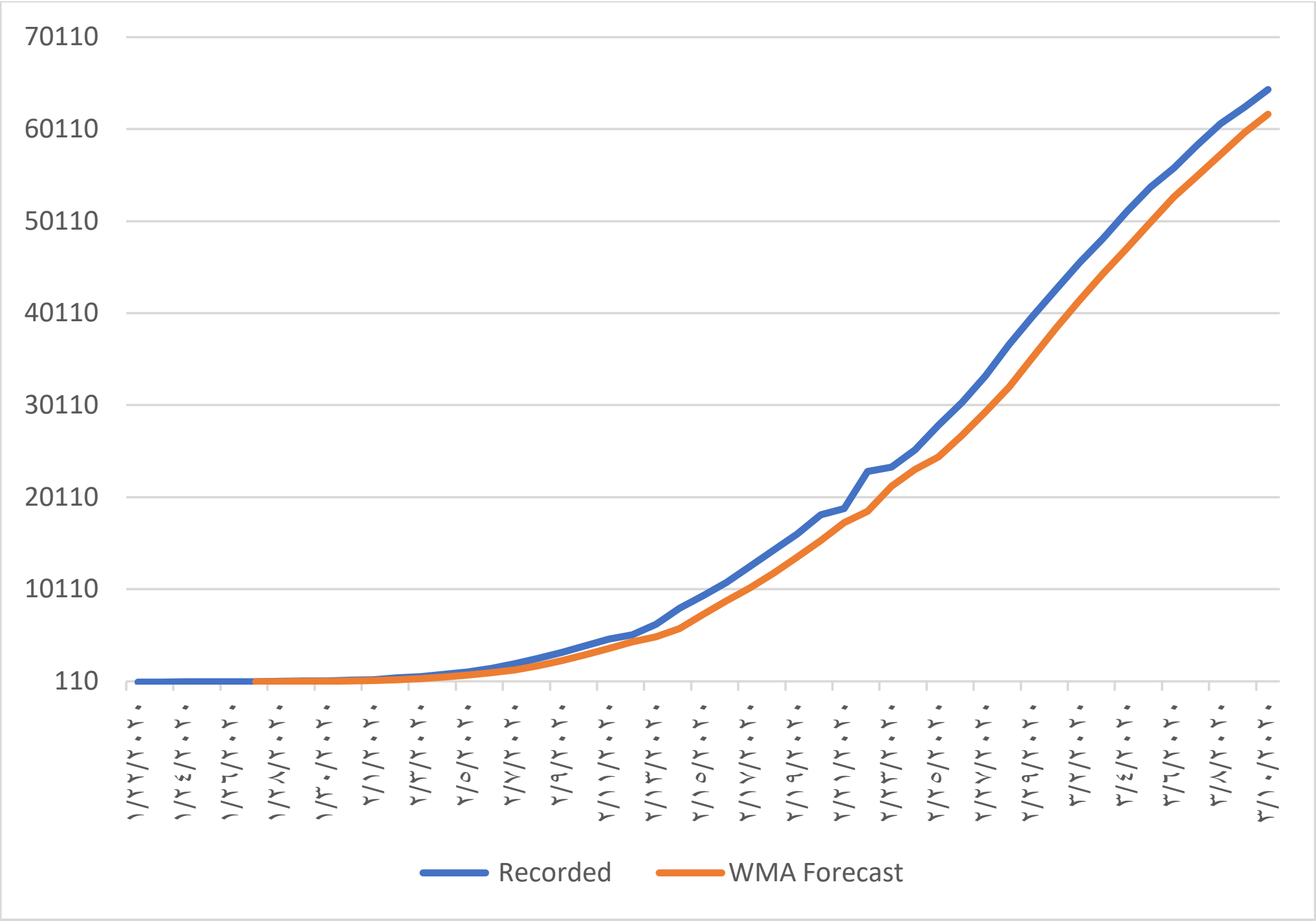

Fig.6 Weighted Moving average (WMA) for forecasting international recovered cases.

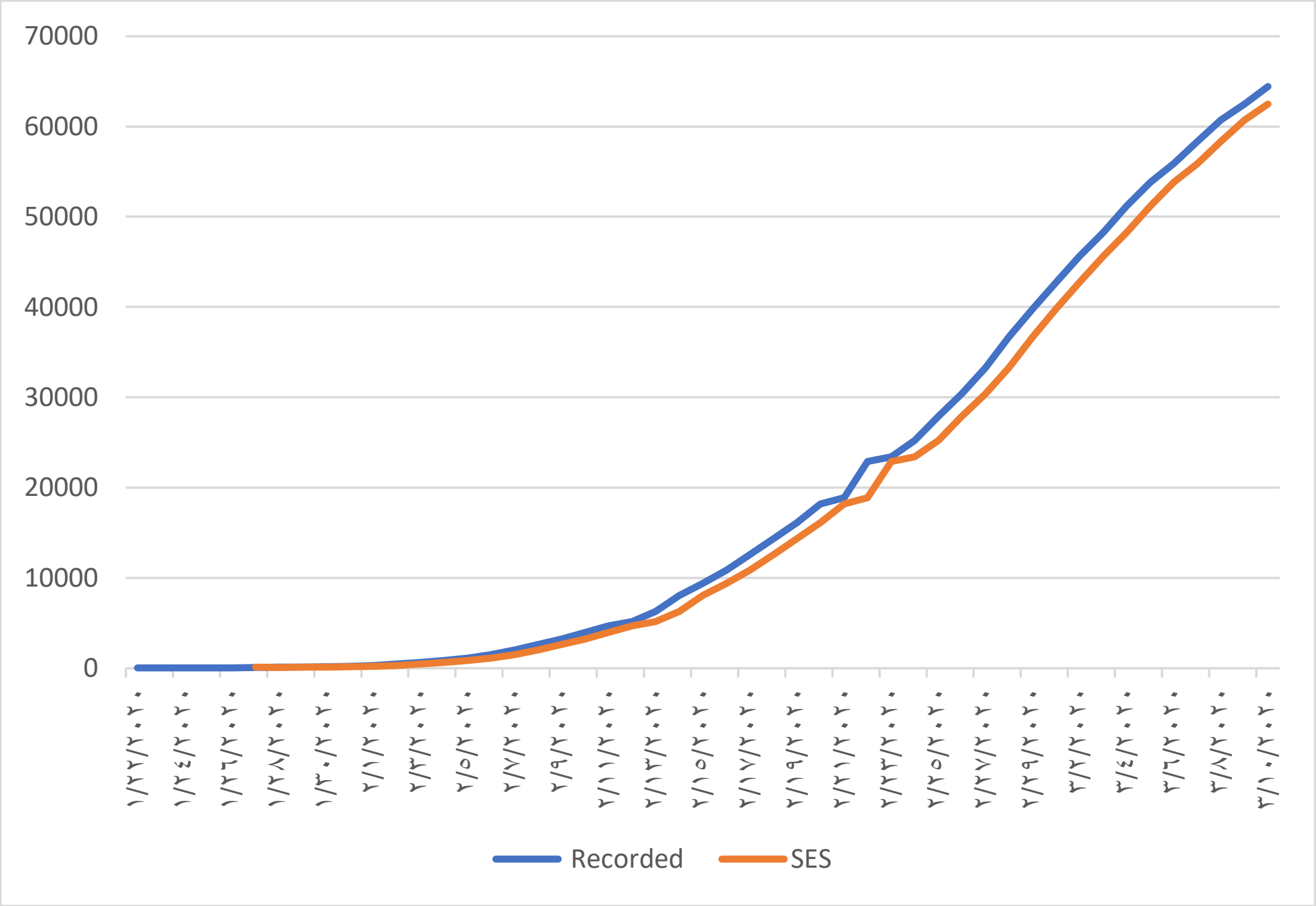

Fig.7 Single Exponential Smoothing (SES) for forecasting recovered cases.

Table.2 Results of forecasting models for international recovered cases.

| Model | MAD | MSE | RMSE | MAPE |
|---|---|---|---|---|
| **A moving average (MA)** | | | | |
| **5 weeks ahead** | 4209.46 | 27558409.86 | 5249.61 | 34.88% |
| **10 weeks ahead** | 2391.04 | 95862422.69 | 9790.94 | 49.48% |
| **Weighted Moving average (WMA)** | 697.39 | 956531.43 | 978.02 | 19.30% |
| **Single Exponential Smoothing (SES)** | 517.54 | 523335.16 | 723.42 | 16.38% |

Third, according to forecasting death cases, Fig.8 shows the moving average (MA) model for forecasting recovered cases for 5 weeks ahead and 10 weeks ahead scales. Fig.9 weighted moving average (WMA) for forecasting death cases where the used weights are 0.0%, 0.0%, 1.5%, 38.1%,60.3%. Fig.10 Single Exponential Smoothing (SES) for forecasting recovered cases. Alpha is optimized based on a genetic algorithm to be equal to 1.00. The three forecasting models have successfully fit the recorded data of the death data. As shown in Table.3, the results of forecasting models for death cases are ranging from 23.72% to on 9.43% MAPE scale. The results indicate that Single Exponential Smoothing (SES) is the most accurate model for forecasting death cases COVID-19 with 82.44, 9685.59, 98.42, and 9.43% for means absolute deviation (MAD), mean square error (MSE), root mean square error (RMSE), means absolute percentage error (MAPE), respectively.

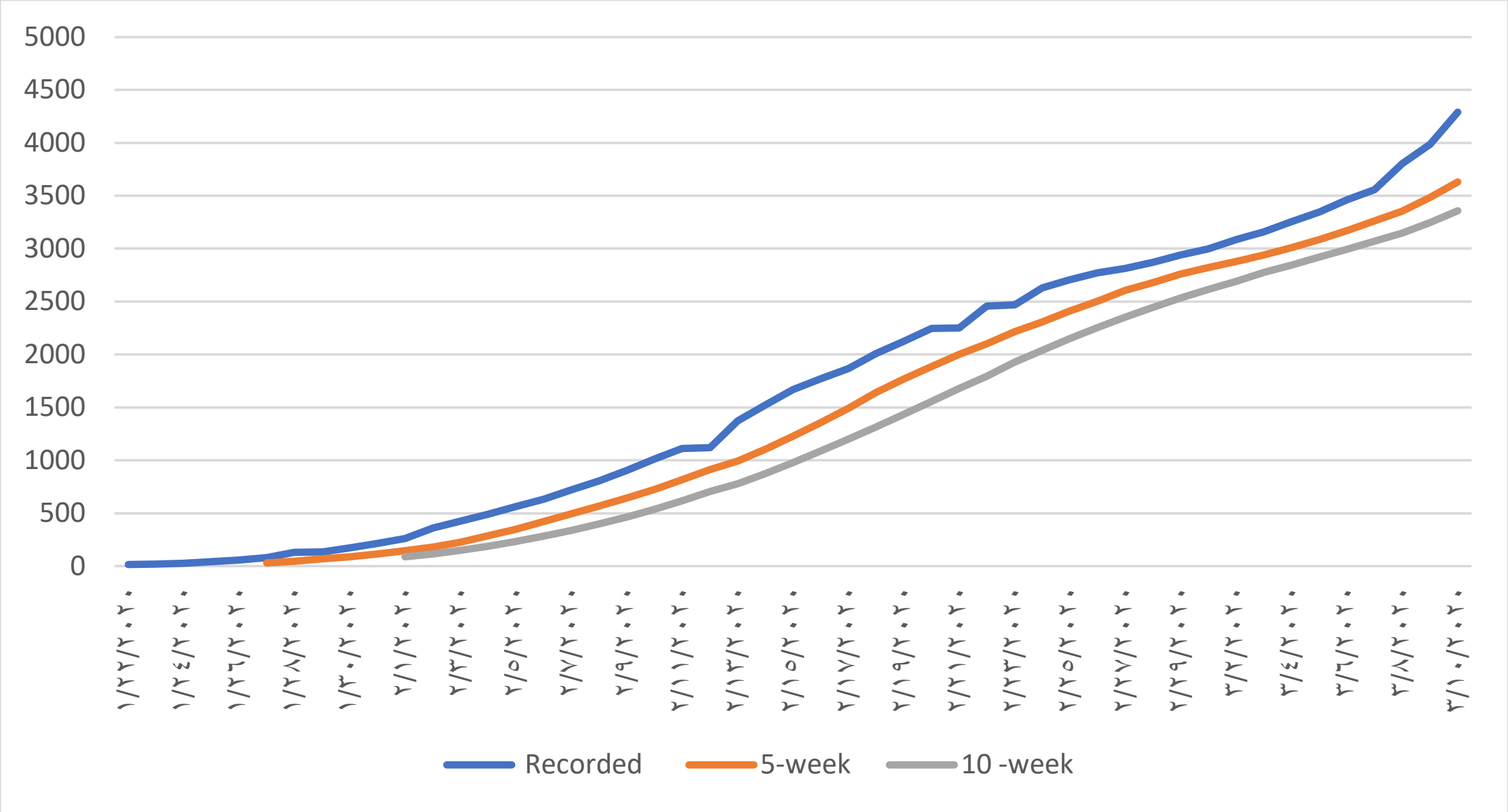

Fig.8 Moving average (MA) for forecasting international death cases.

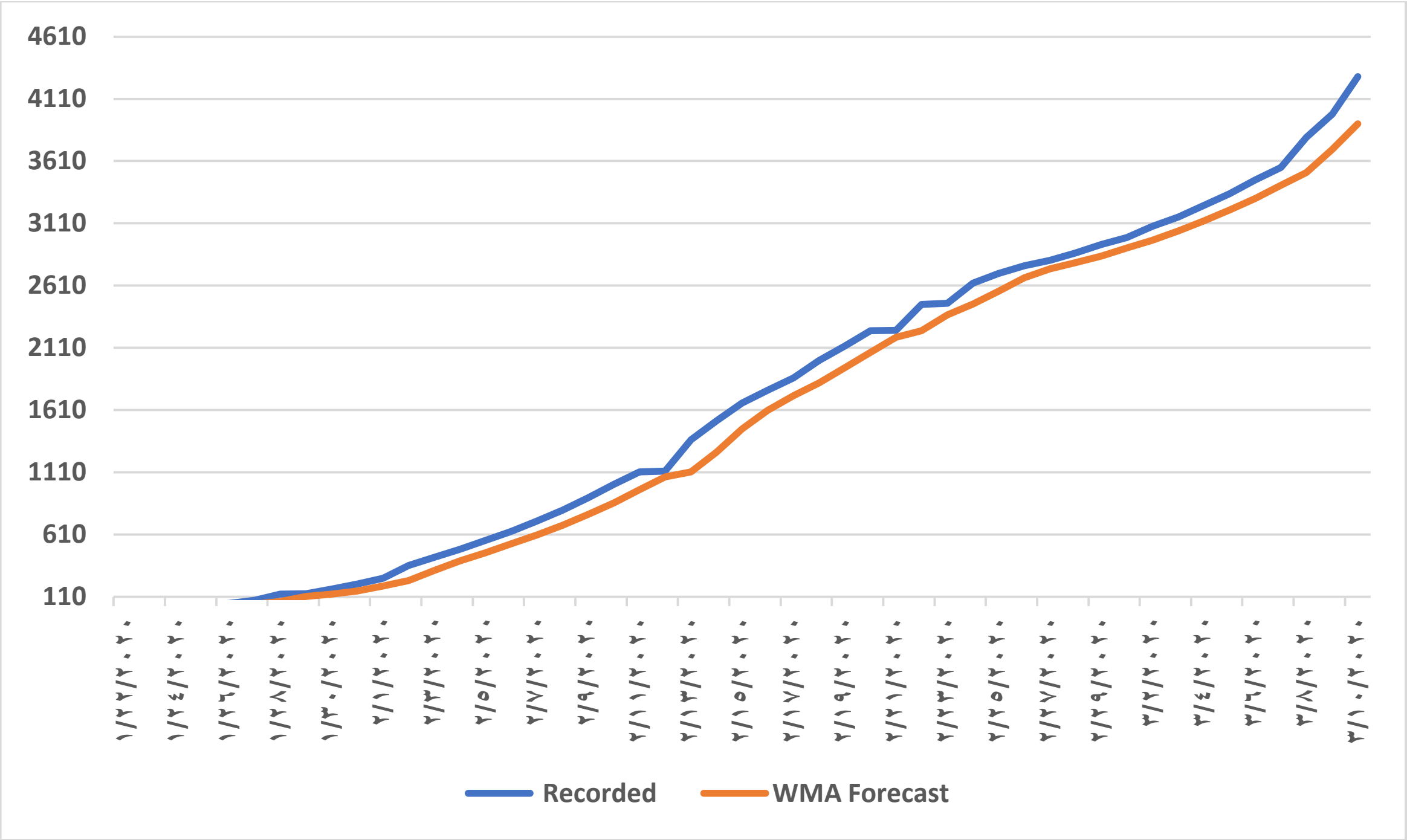

Fig.9 Weighted Moving average (WMA) for forecasting international death cases.

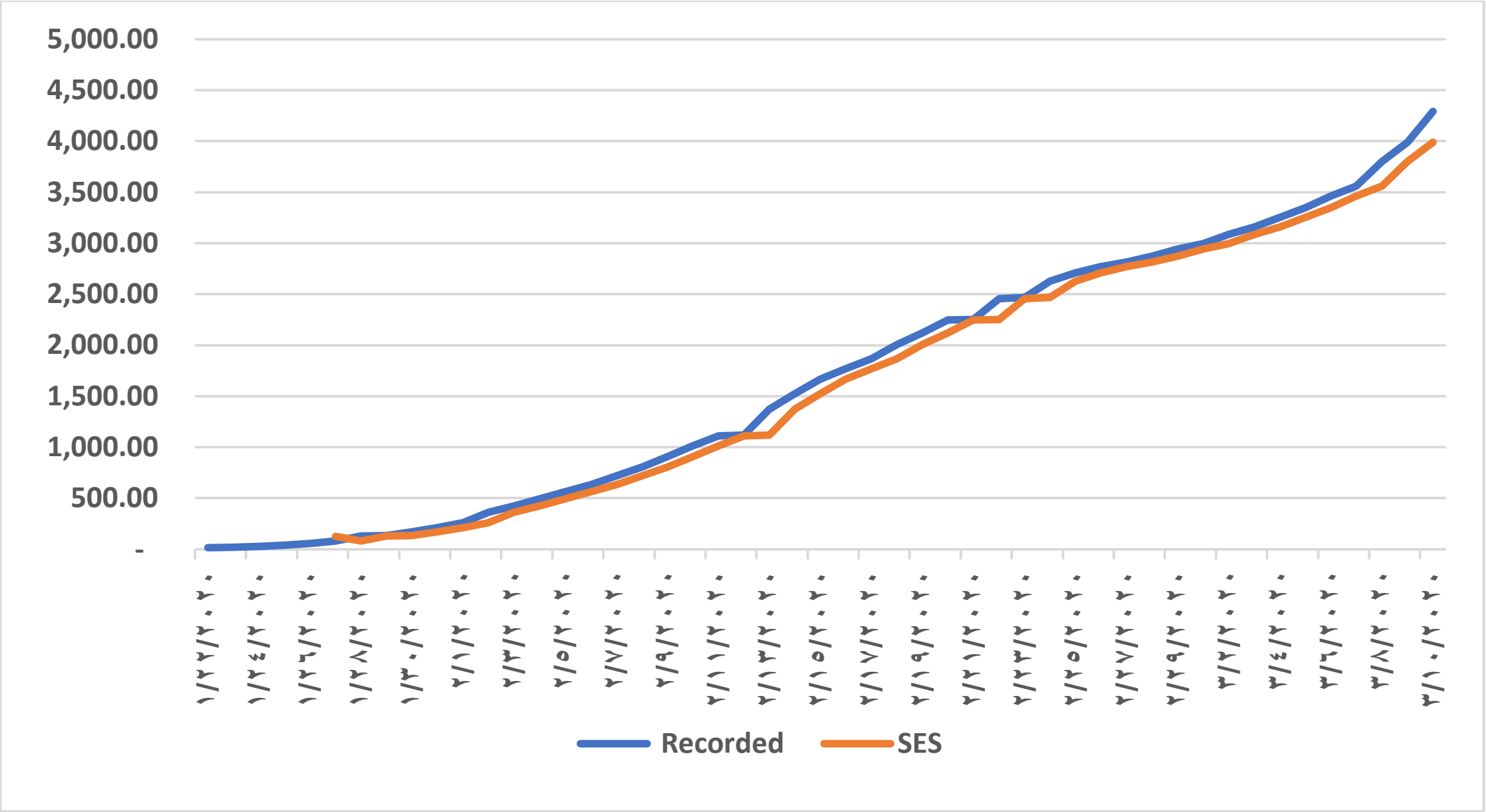

Fig.10 Single Exponential Smoothing (SES) for forecasting international death cases.

Table.3 Results of forecasting models for international death cases.

| Model | MAD | MSE | RMSE | MAPE |
|---|---|---|---|---|
| **A moving average (MA)** | | | | |
| **5 weeks ahead** | 267.30 | 86170.84 | 293.55 | 23.72% |
| **10 weeks ahead** | 432.05 | 278642.61 | 527.87 | 32.93% |
| **Weighted Moving average (WMA)** | 113.37 | 17056.99 | 130.60 | 12.28% |
| **Single Exponential Smoothing (SES)** | 82.44 | 9685.59 | 98.42 | 9.43% |

## 4. Discussion

The results indicate that that Single Exponential Smoothing (SES) is the most accurate model for forecasting confirmed, recovered and death cases COVID-19. The trend analysis with 95% confidence interval shows the followings:

- The number of forecasting confirmed cases will be expected to be increased up to more than 200000 cases around the world in 9th of April 2020 as shown in Fig.11
- The number of forecasting recovered cases will be expected to be increased up to more than 140000 cases around the world on the 9th of April 2020. Accordingly, the recovered ratio will be 96.5% as shown in Fig.12.
- The number of forecasting death cases will be expected to be increased up to more than 7000 cases around the world on the 9th of April 2020. Accordingly, the death ratio will be 3.5% as shown in Fig.13.

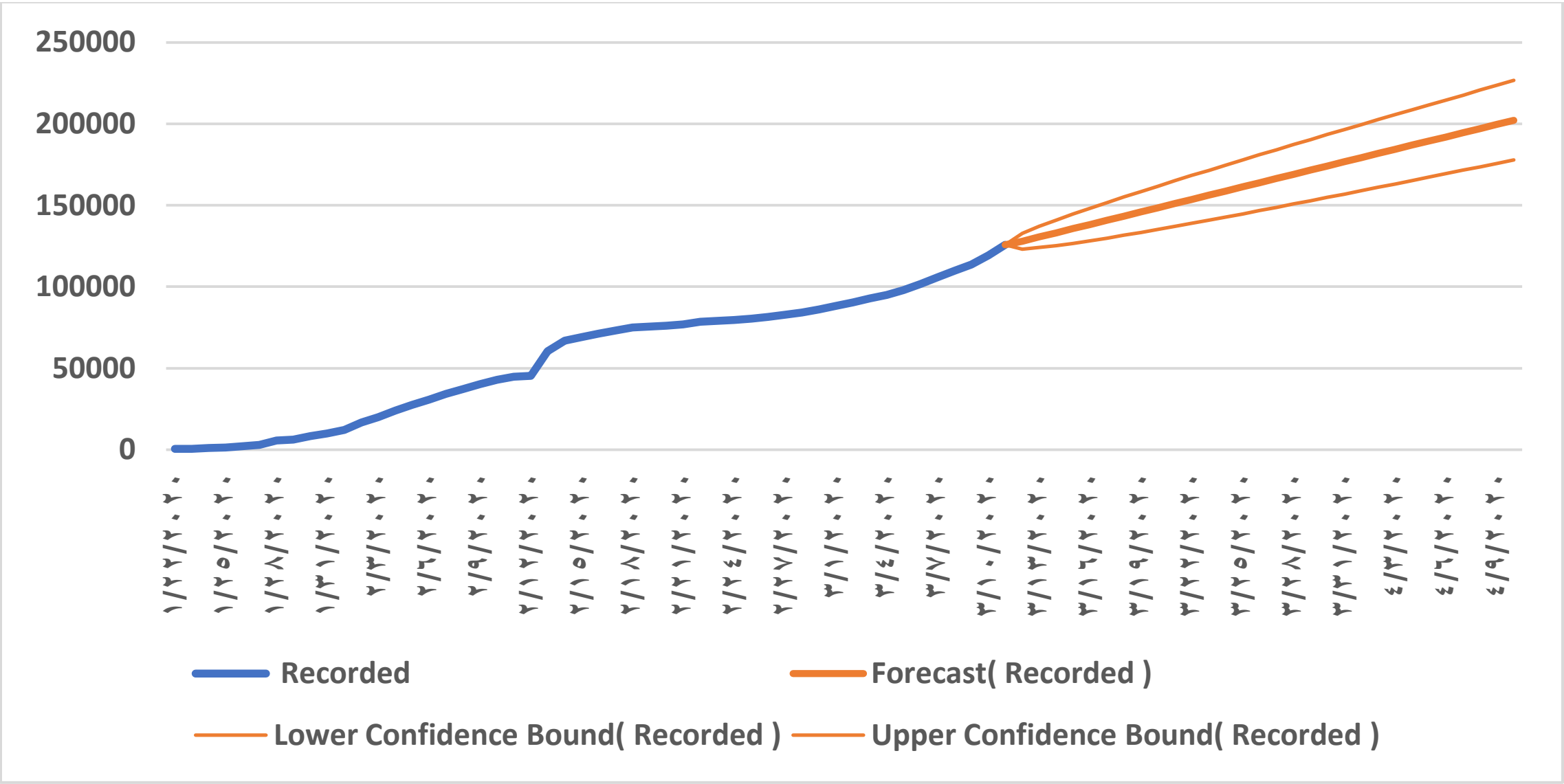

Fig.11 One month ahead trend analysis for confirmed cases.

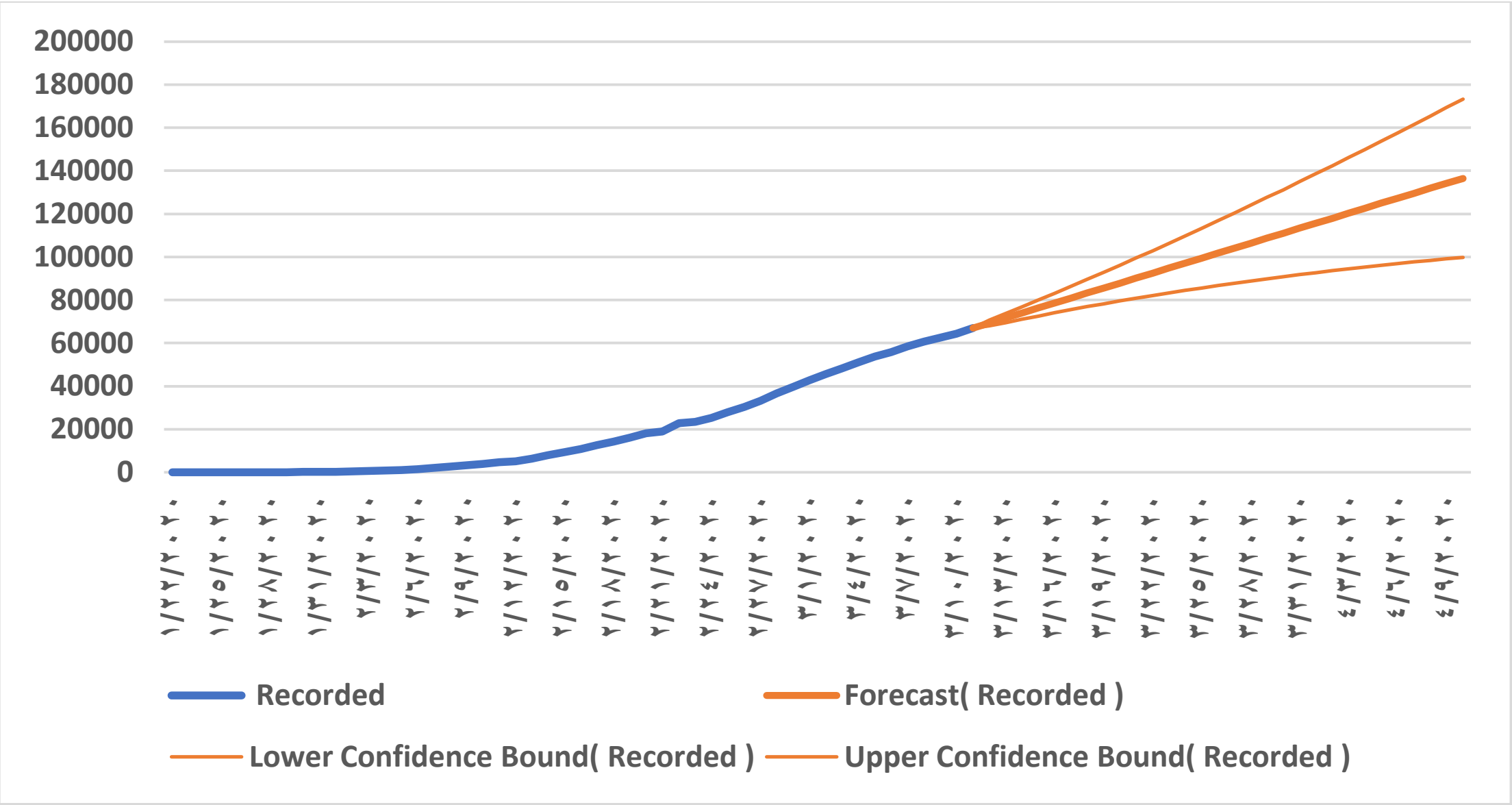

Fig.12 One month ahead trend analysis for forecasting recovered cases.

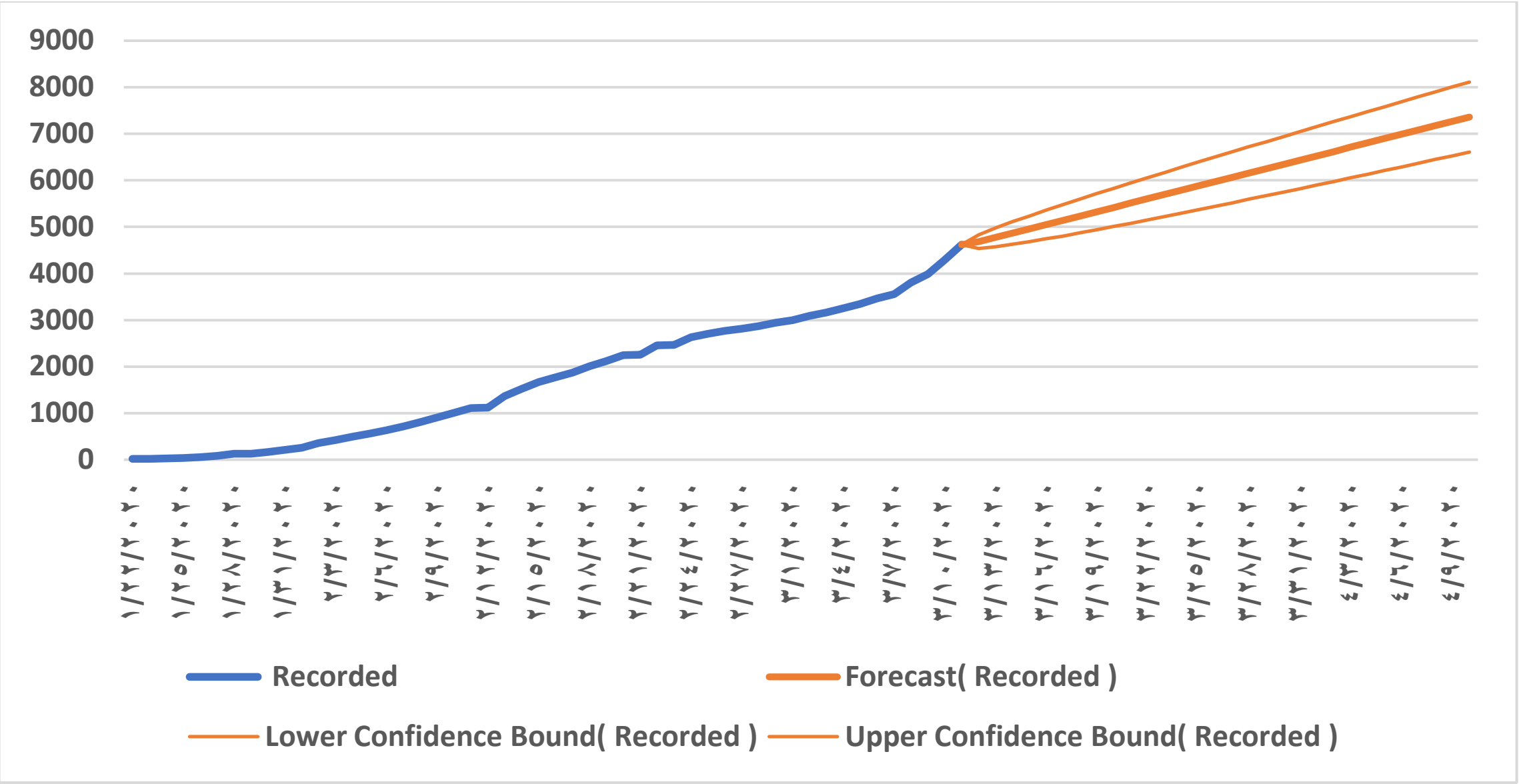

Fig.13 One month ahead trend analysis for death cases.

On the other hand, mathematical models can be formulated to describe and analyze the dynamic of COVID-19 spreading using the data cases outside China mainland. The number of new infected cases and the daily growth rate of conformed cases are proportional to the existing cases. The daily growth rate can be mathematically modeled as the following Equation.

$$\Delta N_d = E*P*N_d \quad (7)$$

Accordingly:

$$N_{d+1} = N_d + \Delta N_d = (1+ E*P)\, N_d \quad (8)$$

Where
$\Delta N_d$: the number of confirmed cases on a given day.
$N_{d+1}$: the expected number of confirmed cases in the next day.
E: average number of people someone infected is exposed to each day.
P: the probability of each exposure becoming an infection.
$N_d$: the number of the coming days (day).
d: the given day.

The growth factor can be calculated as follows:

$$\mathbf{GF} = \frac{\Delta N_d}{\Delta N_{d-1}} \qquad \mathbf{(9)}$$

Where
GF: The growth factor of the confirmed cases of the virus.
$\Delta N_d$: the number of new confirmed cases on a given day.
$\Delta N_{d-1}$: the number of new confirmed cases in the previous day.

Fig.14 displays the growth factor of the confirmed cases outside mainland China. The growth factor was at peak 1.8 on 25 January 2020 and was fluctuating from 23 January 2020 to 14 February 2020. Steady growth exists from 15 February 2020 to 10 March 2020 where the growth factor increased steadily from 1.00 to 1.20

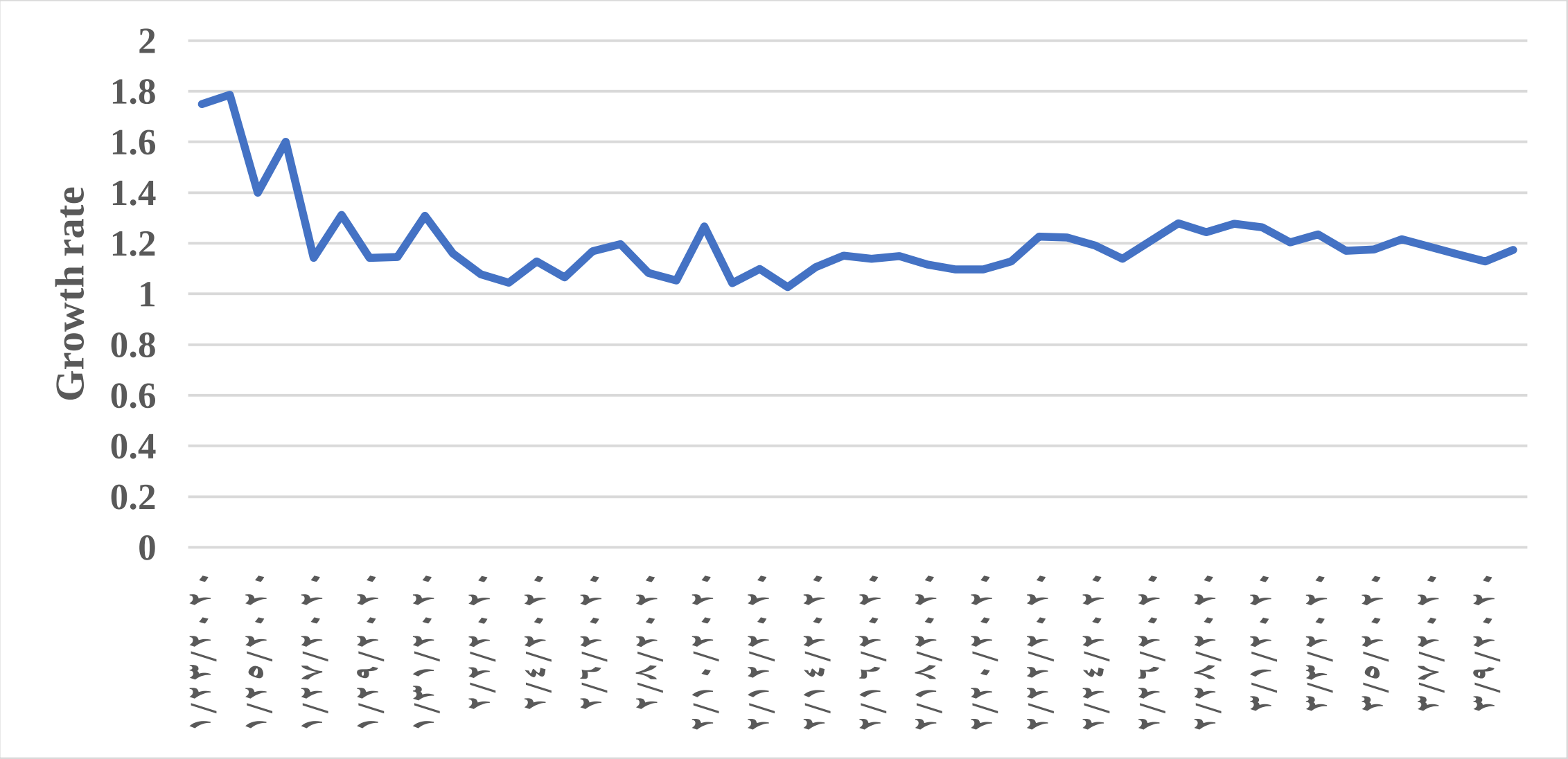

Fig.14 the growth factor (GF) outside mainland China.

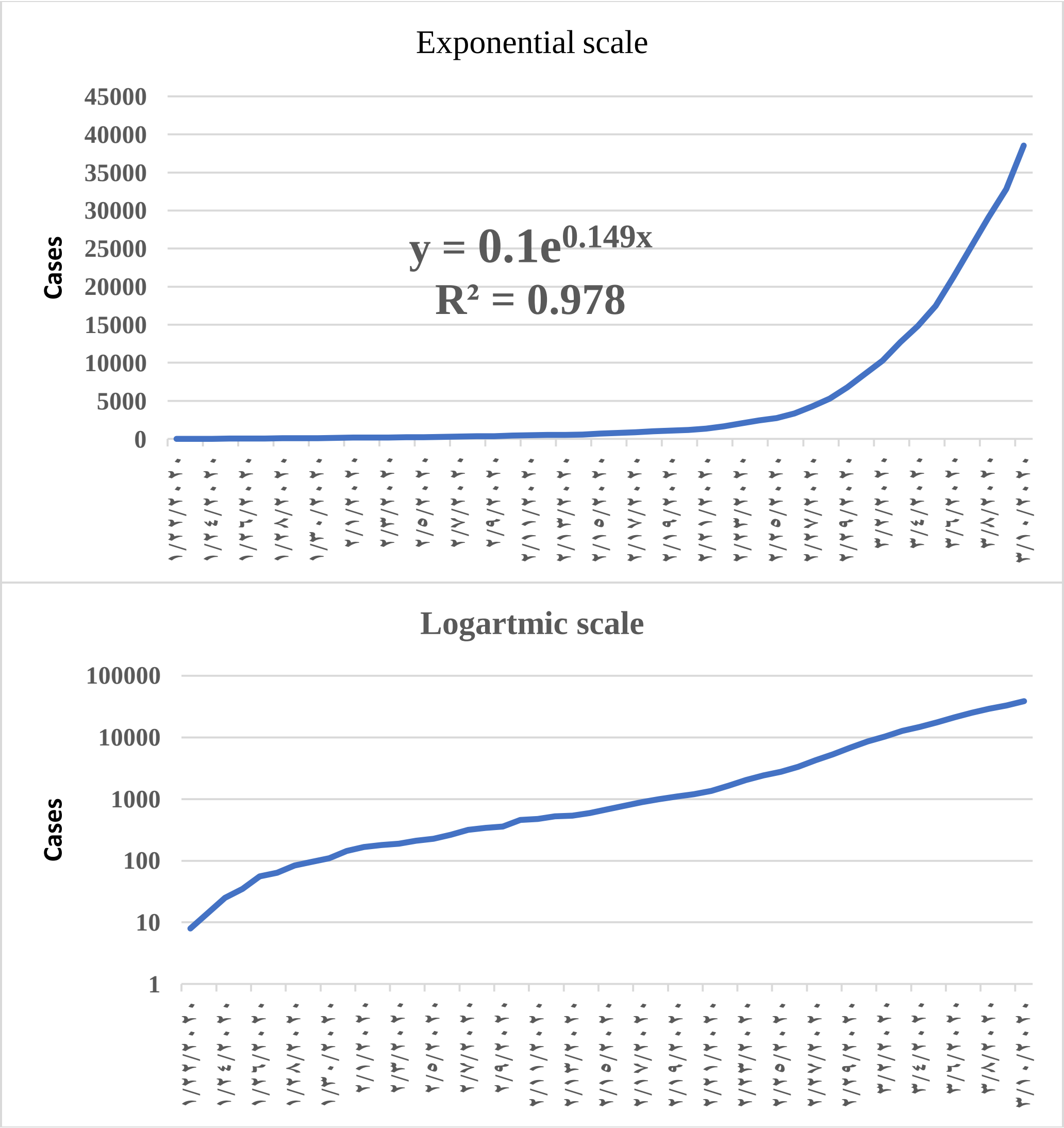

Fig.15 Recorded COVID-19 cases outside mainland China.

The inverse of an exponential function is a logarithmic function. The data were compared on a logarithmic scale to identify exponential growth in the recorded cases. The recorded COVID-19 cases outside mainland China have been plotted as in both exponential scale and in logarithmic scale as illustrated in Fig.15. Based on the exponential chart, the expected growth can be represented by the following exponential function.

$$\boldsymbol{Y = 0.1e^{0.149X}} \qquad \textbf{(10)}$$

Where Y is the expected number of confirmed cases outside mainland China (Cases), and X is the number of the coming days. The accuracy of the exponential formula is 97.85% on the $R^2$ scale. Based on the exponential formula, on average, the number of confirmed cases outside mainland China has doubled 10 folds every 16 days.

South Korea had 6593 cases and New Zealand had 64 cases where this means that the cases of South Korea are 100 folds the cases of New Zealand where the first case in New Zealand was discovered a month behind the first discovered case in South Korea. Thus, only one month is sufficient to double the cases to 100 folds. The logarithmic scale in Fig.15 illustrates the following:

- The logarithmic scale chart indicates that during 6 days from 22/1/2020 to 28/1/2020, the cases jumped from 10 to 100 cases outside China.
- During 20 days from 28/1/2020 to 18/2/2020, the cases jumped from 100 to 1000 cases outside China.
- During 13 days from 18/2/2020 to 1/3/2020, the cases jumped from 1000 to 10000 cases outside China.

In the future, the trend of the logarithmic scale chart shows that the total number of confirmed cases outside China is expected to be one million, 10 million, 100 million, one billion in the next 30, 47, 64, and 81 days respectively. Mathematically, it can be formulated as follows:

$$\mathbf{N_d = (1+ E*P)^d N_0} \qquad \mathbf{(11)}$$

Where

$N_d$: the expected number of the confirmed cases in the future.

E: average number of people someone infected is exposed to each day.

P: the probability of each exposure becoming an infection.

$N_0$: the initial number of cases at a given time.

d: the number of days between the given time and the future time.

Therefore, the number of confirmed cases in a given day ($N_d$) will decrease when E or P decreases. The restrictions on travel and public gatherings, and closing of schools, universities, and workplaces ("Social Distancing") is the current possible way to decrease E and P.

Using the date of the first confirmed coronavirus case in each country as its day 0, data analysis has been conducted to predict the trajectory of the 2019 COVID-19 outbreak on a country-by-country basis for each day as shown in Fig.16. As a result, the number of coronavirus cases compounds at more than 25% per day in countries that do not mandate social distancing. This behavior is consistent across Iran, Italy, Spain, France, and the UK, each measured independently from the date of the first case in that country. Moreover, at a density of approximately 9.7 coronavirus cases per 100,000 inhabitants, Italy mandated on 8$^{th}$ March 2020 a complete regional lock-down, with movement restrictions applying to about 16 million people (25% of the population of Italy). The following day, on 9$^{th}$ March 2020, the lock-down was extended to the entire country (100% of the population of Italy).

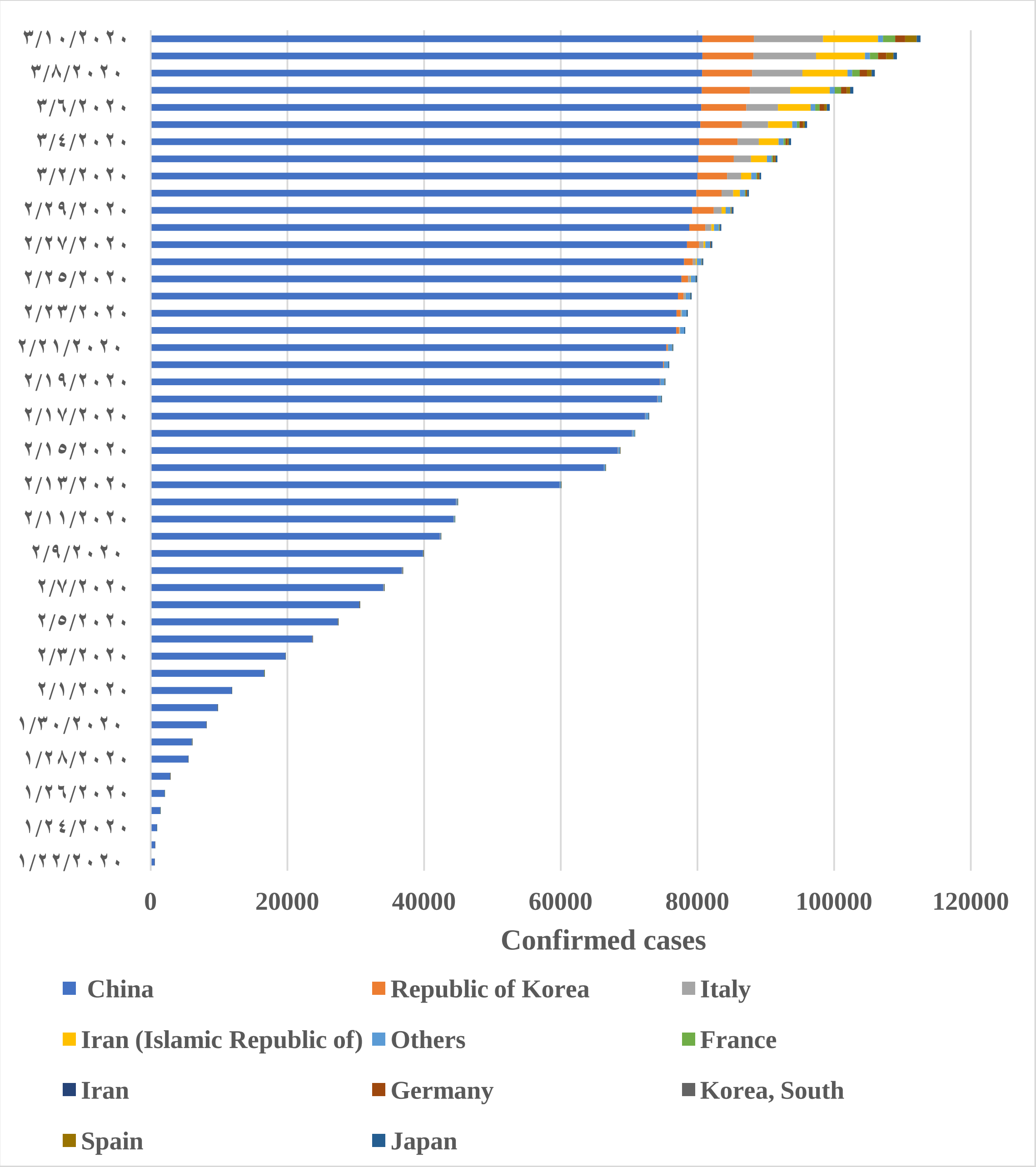

Fig.16 COVID-19 outbreak on a country-by-country basis for each day.

For example, in Egypt, the country has not yet closed its borders. The current population of Egypt is 102,334,404 inhabitants where the confirmed cases ratio is 1.06 case / one million inhabitants**.** Up to data the death rate is approximately 2% in Egypt. As shown in Fig.17, No Covid-19 exists till 14th February 2020 when the first confirmed case was recorded. The second case was recorded om 1st March 2020 and the third cases was in 5th March 2020. However, the confirmed cases have exponentially doubled to 15,49, 60, and 109 on 6th March 2020, 8th March 2020, 11th March 2020, and 14th March 2020, respectively. Based on the Logarithmic scale, the cases doubled approximately 10-folds from 15 to 109 during only 8 days with two death cases and 27 recovered cases.

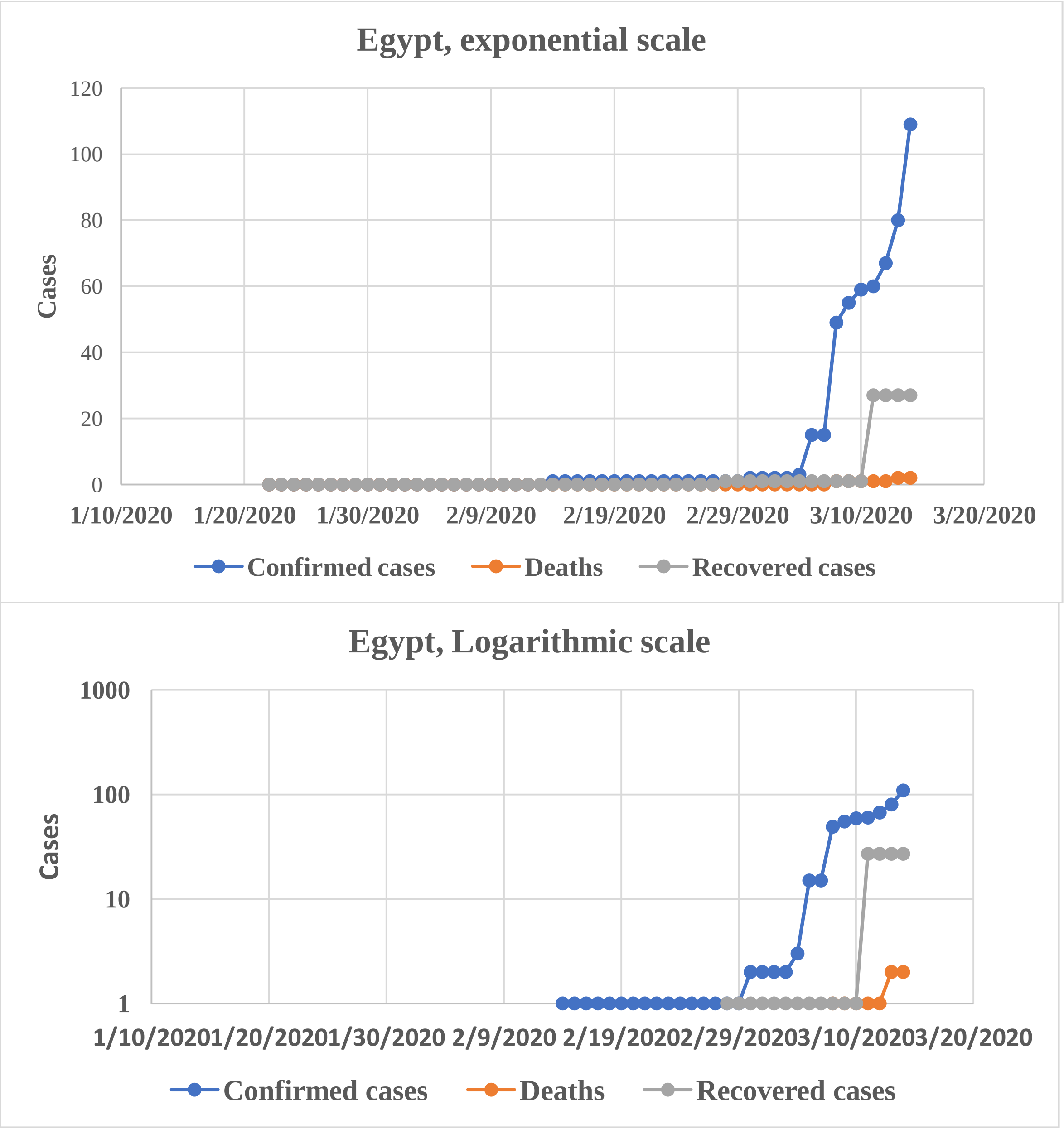

Fig.17 COVID-19 outbreak in Egypt for each day.

As shown in Fig.18, COVID-19 confirmed cases change rate in Egypt steady increases from 12% to 36% in 9$^{th}$ March 2020 to 14$^{th}$ March 2020, respectively. However, this change rage has dramatically increased in to 500% and 320% in 6$^{th}$ March 2020 and 8$^{th}$ March 2020, respectively. The forecasting model with 95% confidence interval expects that the number of confirmed cases in Egypt will be doubled up to 4-folds in the next month without any sudden raise in the infection change rage. Therefore, the forecasting models and data strongly suggest that the number of coronavirus cases grows exponentially in Egypt as displayed in Fig.19. Therefore, Egypt and all other countries must immediately mandate quarantines, restrictions on travel and public gatherings, and closing of schools, universities, and workplaces to achieve required safe social distancing among the humans.

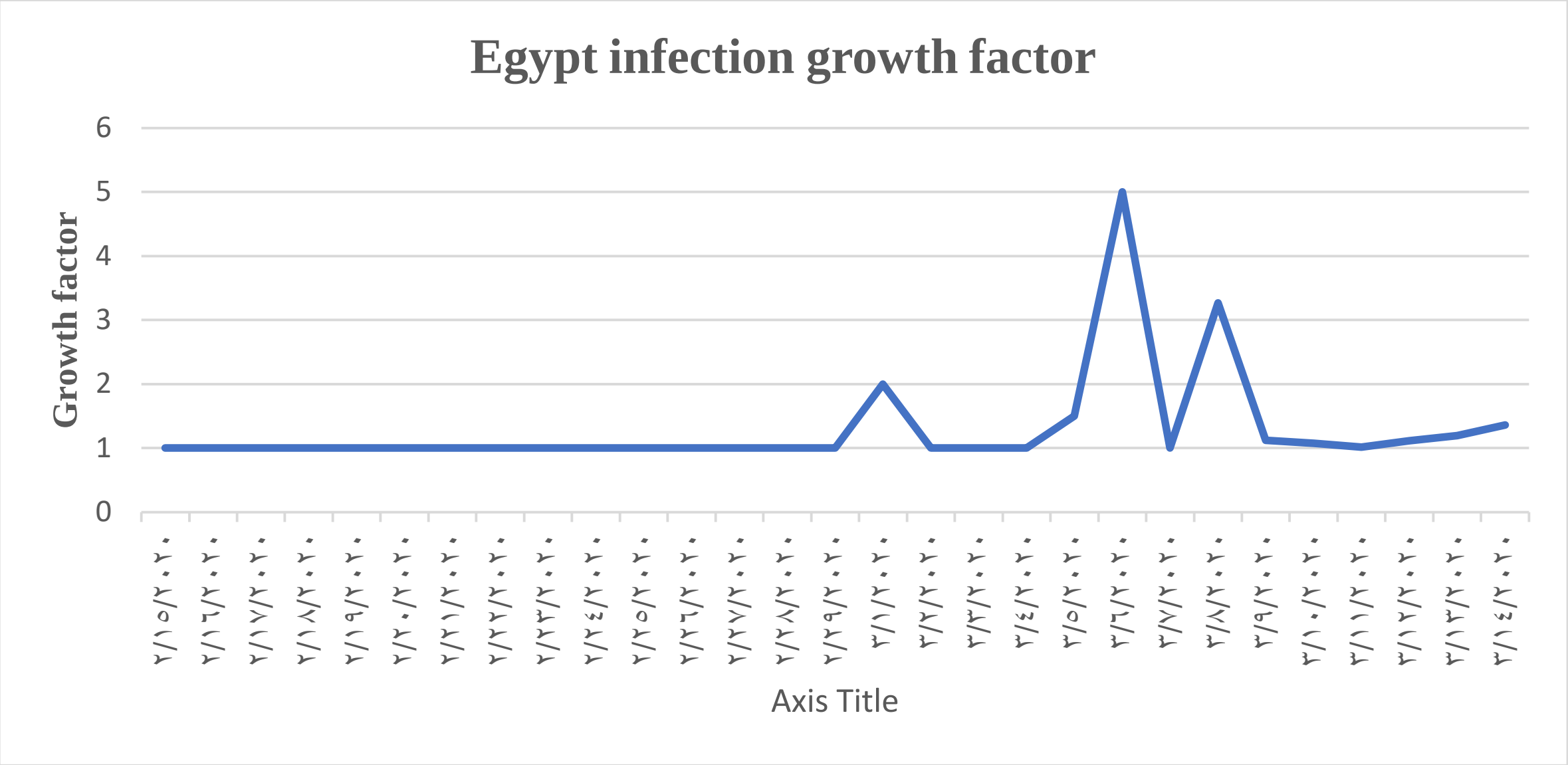

Fig.18 COVID-19 confirmed cases growth factor in Egypt.

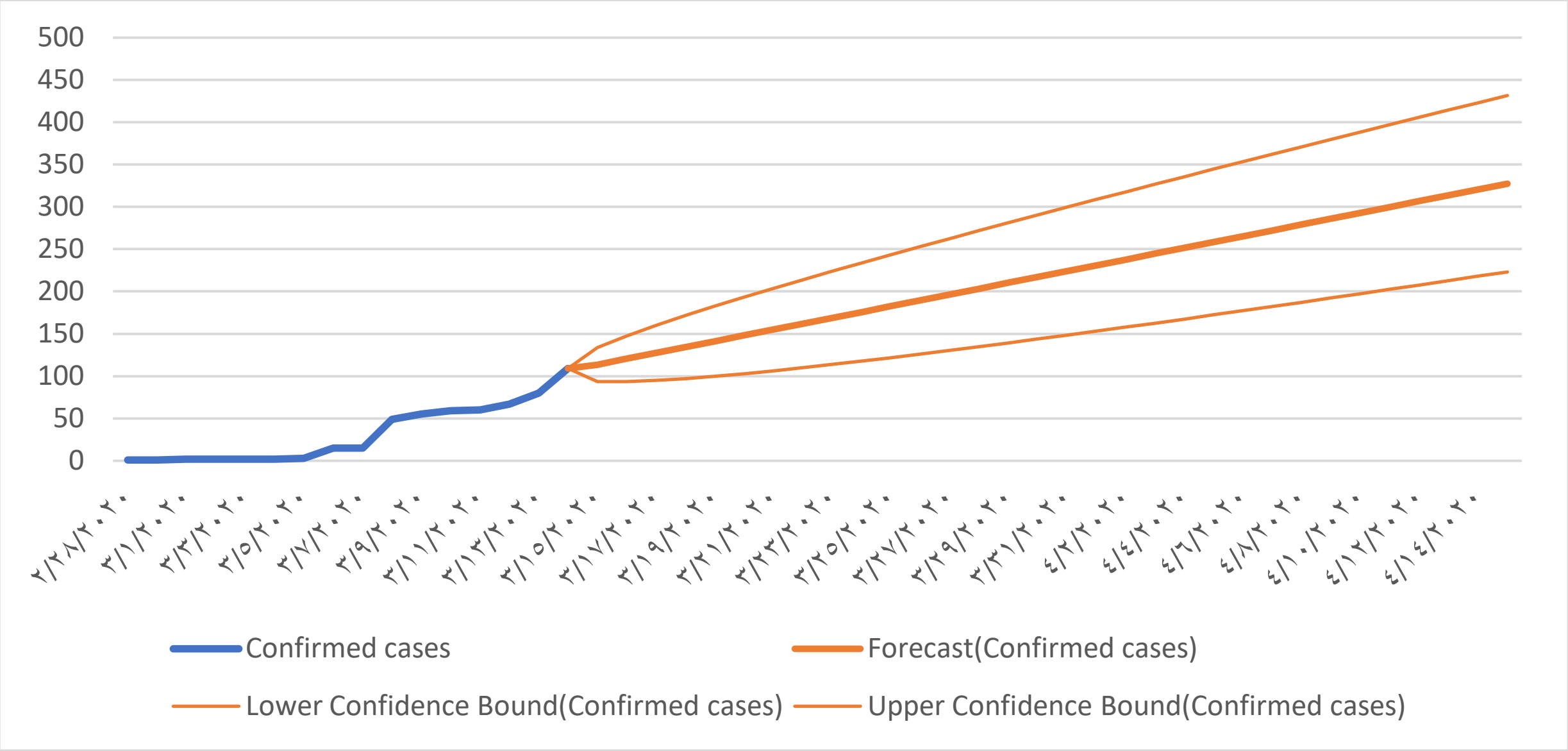

Fig.19 Forecasting confirmed cases in Egypt.

Another example, by March 30, 2020, United States (US) density will be expected to reach the Italian threshold of 9.7 US coronavirus cases per 100,000 US inhabitants. This threshold density is equal to the density of cases when Italy implemented a regional lock-down that affected 25% of its population. by April 9, 2020, the density could reach 87 US coronavirus cases per 100,000 US inhabitants, unless further social controls are adopted. A total of 87 cases per 100,000 inhabitants is computed as the threshold density at which all currently unoccupied intensive care unit (ICU) beds would be occupied by a US person who is infected with coronavirus and has severe, life-threatening symptoms. In other words, no regional beds to house patients who are severely ill.

The following insights can be developed based on the trend analysis and the mathematical formulation of the spreading of COVID-19:

- The 2019–2020 coronavirus outbreak is at an inflection point.
- The data strongly suggests that the number of coronavirus cases grows exponentially in countries that do not mandate quarantines, restrictions on travel and public gatherings, and closing of schools, universities, and workplaces ("Social Distancing"). The most striking differences (between China, Hong Kong, and Japan versus the rest of the world) confirm the hypothesis of person-to-person transmission as the driver of exponential growth since the countermeasures they have taken specifically limit this and do nothing to change the detection rate.
- The number of coronavirus cases compounds at more than 25% per day in countries that do not mandate social distancing. This behavior is consistent across Iran, Italy, Spain, France, and the UK.
- Log of several cases per capita for various countries as a function of time. Without social isolation measures, growth is 25-35% per day. Lock-downs do bring down the rate of new infections.
- The highly precise exponential growth strongly suggests that the growth in these cases is due to an underlying biological phenomenon (e.g. virus transmission) rather than due to an increase in the availability of tests or the number of tests performed.

## 5. Conclusions

Coronaviruses such as COVID-19 is declared as an international epidemic. Based on this study analysis and modeling, this research recommends that all world countries must mandate substantially more invasive quarantines, restrictions on travel and public gatherings, and closing of schools, universities, and workplaces ("Social Distancing") in the near term when ICU beds are unavailable and patient deaths begin to rise precipitously. This will affect a significantly larger percentage of the world population than is currently affected by voluntary social distancing, including younger people in the workforce who might be asymptomatic but who are forced to work from home and curtail spending and travel. The number of people affected in the world will increase dramatically within the next three weeks.

**Acknowledgment:** We would like to thank the scientific research group in Egypt (SRGE) for its support of this study.